\documentclass[epj]{svjour}
\usepackage{graphics}
\usepackage{longtable}
\usepackage{amsfonts}
\usepackage{amsmath}
\usepackage{amssymb}
\usepackage{kantlipsum,widetext}
\usepackage{dcolumn}
\usepackage{epsfig}
\usepackage{graphicx}   
\usepackage[utf8]{inputenc}
\usepackage{latexsym}
\usepackage{color}
\begin{document}
\title{Brane inflation and the robustness of the Starobinsky inflationary model}
\author{S. Santos da Costa\inst{1} \and M. Benetti\inst{2,3,4} \and R.M.P. Neves\inst{5}  \and F. A. Brito\inst{5,6} \and R.  Silva\inst{7,8} \and  J.  Alcaniz\inst{1} % etc
}                     % Do not remove
%
%\offprints{simonycosta@on.br}          % Insert a name or remove this line
\mail{simonycosta@on.br}
\institute{Departamento de Astronomia, Observat\'orio Nacional, 20921-400, Rio de Janeiro, RJ, Brasil  \\ \email{simonycosta@on.br}; \email{alcaniz@on.br}
\and Dipartimento di Fisica ‘E. Pancini’, Universit\'a di Napoli ‘Federico II’, Compl.
Univ. di Monte S. Angelo, Edificio G, Via Cinthia, I-80126, Napoli, Italy \\ \email{micolbenetti@unina.it}
\and Istituto Nazionale di Fisica Nucleare (INFN) Sez. di Napoli, Compl. Univ. di Monte
S. Angelo, Edificio G, Via Cinthia, I-80126, Napoli, Italy 
\and Scuola Superiore Meridionale, Universit\`{a} di Napoli ``Federico II'', Largo San Marcellino 10, 80138 Napoli, Italy 
\and Unidade Acad\^emica de F\'isica, Universidade Federal de Campina Grande, 58109-970 Campina Grande, PB, Brasil  \\ \email{fabrito@df.ufcg.edu.br}  \and Departamento de F\'isica, Universidade Federal da Para\'iba,  58051-970, Jo\~ao Pessoa, PB, Brasil \\ \email{raissapimentel.ns@gmail.com}
\and Departamento de F\'isica, Universidade Federal do Rio Grande do Norte, 59072-970, Natal, RN, Brasil \\ \email{raimundosilva@dfte.ufrn.br} 
\and Departamento de F\'isica, Universidade do Estado do Rio Grande do Norte, Mossor\'o, 59610-210, Brasil 
}

%\institute{Insert the first address here \and the second here}
%
\date{Received: date / Revised version: date}
\abstract{The first inflationary model conceived was the one proposed by Starobinsky which includes an additional term quadratic in the Ricci-scalar $R$ in the Einstein-Hilbert action. The model is now considered a target for several future cosmic microwave background experiments given its compatibility with current observational data. In this paper, we analyze the robustness of the Starobinsky inflation by inserting it into a generalized scenario based on a $\beta$-Starobinsky inflaton potential, which is motivated through brane inflation. In the Einstein frame, the generalized model recovers the original model for $\beta = 0$ whereas $\forall \beta \neq 0$ represents an extended class of models that admits a wider range of solutions. We investigate limits on $\beta$ from current cosmic microwave background and baryonic acoustic oscillation data and find that only a small deviation from the original scenario is allowed, $\beta=-0.08 \pm 0.12$ ($68\%$ C.L.), which is fully compatible with zero and confirms the robustness of the Starobinsky inflationary model in light of current observations. 
\PACS{
      {PACS-key}{discribing text of that key}   \and
      {PACS-key}{discribing text of that key}
     } % end of PACS codes
} %end of abstract
\maketitle
\section{Introduction}
\label{intro}

The inflationary framework yields a viable explanation for some problems of the Big Bang cosmology, as well as for the process of growth of the primordial cosmological perturbations which produced the observed large-scale structures and temperature fluctuations in the Cosmic Microwave Background (CMB). The simplest models of inflation involve a single scalar field $\phi$ slowly rolling down its potential $V(\phi)$, which generates primordial scalar perturbations with a nearly scale-invariant power spectrum~\cite{Mukhanov:2005sc,weinberg2008cosmology} (see also \cite{Senatore:2016aui} for a recent review). The recent CMB observations~\cite{Planck2015,Aghanim:2018eyx} have not only confirmed this framework but also allowed to test the observational viability of a number of inflationary models (see e.g. \cite{Martin_2014}). 

Although the majority of models of inflation involve scalar fields, the very first model proposed was driven by quantum corrections to the Einstein-Hilbert Lagrangian~\cite{Starobinsky:1980te} (usually called Starobinsky or $R^2$ inflation), i.e.,
\begin{equation} \label{eh}
S = \frac{\rm{M_{Pl}^2}}{2}\int{d^4x\sqrt{-g}\left(R + \frac{R^2}{\mu^2}\right)}\;,
\end{equation}
which includes a quadratic term of Ricci scalar, $R^2$, that dominates the Lagrangian density during the primordial universe -- in the above expression, $\rm{M_{Pl}}$ is the Planck Mass and $\mu$ is a given mass scale. The equivalence between Einstein and Jordan frames through a conformal transformation of the metric allows to deal with an inflaton potential of type 
\begin{eqnarray} 
 V(\phi)=V_0 \left[1 - \exp\left(-\sqrt{\frac{2}{3}}\frac{\phi}{\rm{M_{Pl}}}\right)\right]^2, \label{eq:pot_sta}
\end{eqnarray}
where $V_0$ is the amplitude of the potential. The above expression is the equivalent of the $R^2$ contribution to the Lagrangian density (see \cite{Ketov:2019toi} and references therein for more details) and describes a class of potentials that obeys the slow-roll approximation, necessary for inflation to happen and produce the in-homogeneity pattern observed in the CMB data.

From the theoretical side, recent investigations have shown that inflationary potentials of several unrelated inflationary models coincide, leading to identical predictions for the slow-roll parameters $n_s$ and $r$, which well fit observational data~\cite{Aghanim:2018eyx}. The original Starobinsky model, for instance, is a particular case whose potential emerges in i) the Higgs model with a non-minimal coupling to gravity, $\xi \phi^2 R - \frac{\lambda}{4}(\phi^2 - v^2)^2$,  for $\xi < 0$, in the limit $1 + \xi v^2$ ~\cite{Linde:2011nh}, ii) as a simple conformally invariant theory with spontaneous symmetry breaking, in the context of superconformal theory and supergravity~\cite{Kallosh:2013xya,Kehagias:2013mya}, and iii) in the large field regime of a Superconformal D-Term Inflation~\cite{Buchmuller:2013zfa}. {More recently, it was shown that a quadratic term of Ricci scalar in the Lagrangian also arises from a wide family of string models by using the Noether Symmetry Approach~\cite{Benetti:2019smr,Capozziello:2015hra}.}
%In addition, the standard Starobinsky model, i.e. the one with the quadratic term of Ricci scalar in the Lagrangian, arises from a wide family of string models by using the Noether Symmetry Approach~\cite{Benetti:2019smr,Capozziello:2015hra}.}% Because f(R) gravity can be understood as a sector of Lovelock gravity which coincides with Einstein gravity only in four-dimensions, but at higher spacetime dimensions resembles string theory inspired models of gravity it seems natural to make a further connection of the Starobinsky model and its modifications as being part of the theories that can be found in fundamental setups such as in the realm of string/brane theories.

From the observational viewpoint, analyses of different classes of inflationary models using current CMB data have shown that the Starobinsky model provides an excellent fit to the data~\cite{Martin_2014}, being now considered as a ``target" model for some planned CMB experiments (see e.g. \cite{Abazajian:2016yjj,Suzuki:2018cuy,Ade:2018sbj}). The model predicts a spectral index $n_s \simeq 0.96$ with a small spectral running and also a small amount of gravitational waves. Given its compatibility with current observational data, extensions of the Starobinsky model have been investigated. For instance, a simple extension including an extra scalar field was studied in \cite{vandeBruck:2015xpa}. Furthermore, attempts in the context of higher derivative theories of the type $R^{2p}$ and other extensions of the Starobinsky $R^2$ model were also considered in \cite{Renzi:2019ewp,Sebastiani:2013eqa,Myrzakulov:2014hca}. The analysis performed in \cite{Renzi:2019ewp} considered the Einstein frame in searching for deviations from the benchmark value of the tensor amplitude for the case with $p \simeq 1$, which recovers the Starobinsky model. It was found that the original Starobinsky model provides an excellent fit to the CMB data, despite the fact that uncertainties on $n_s$ may modify the expected value of $r$. 

Our goal in this paper is to investigate the robustness of the Starobinsky scenario in light of current observational data. In principle, to check the robustness or validity of a theory or model, it is important to insert it into a more general framework. The general framework will be based on the derivation of a generalized inflaton potential -- from now on $\beta$-Starobinsky (See eq. (7)), which depends on a parameter $\beta$ and extends the potential (2). As it will be discussed later, constraints on the parameter $\beta$ quantify directly the allowed deviations from the original model and, therefore, its robustness with respect to increase in the number of degrees of freedom and also to the observational data. Inflationary models driven by generalized exponential potentials have also been investigated in \cite{Alcaniz:2006nu,Santos:2017alg,Gron:2018rtj}.

We organize this paper as follows. In sect. \ref{braneinflation}, we present the route in order to deduce the $\beta$-Starobinsky inflaton potential in the brane inflation context. In sect. \ref{betastarobinskyinflation},  we discuss the main features of the potential given by eq.\eqref{eq:pot_ext} through a slow-roll analysis and compare its theoretical predictions with the latest results of the Planck Collaboration. Section \ref{method} presents the method employed to calculate the theoretical predictions for the amplitude of fluctuations of the CMB temperature and the statistical analyses performed using the current CMB data. A discussion of the main results of our analysis is shown in sect. \ref{results}. We present our conclusions in sect. \ref{conclusions}.

\section{$\beta$-Starobinsky potential from brane inflation}
\label{braneinflation}

Let us discuss a route based on the brane inflation providing the $\beta$-Starobinsky inflaton potential. In ref.~\cite{Santos:2017alg} we previously obtained an induced four-dimensional $\beta$-inflaton potential in a brane inflation scenario given by the general form
\begin{equation}
V_{\rm eff}(L)=A_0(1-c_1 L)^{\frac{1}{\lambda c_1}}+\frac12\sigma,
\end{equation}
where $L$ is the brane position along the fifth dimension in relation to the origin, $\sigma$ is the brane tension, $c_1$ and $\lambda$ are the parameters of the five-dimensional theory. In four dimensions we can interpret $L$ as the inflaton field as $\phi=M_{Pl}^2 L$.

By redefining parameters as $\lambda c_1\to\beta$, $L\to  \lambda^2\phi$, we find%\footnote{We have changed the sign of L, since it only indicates the position of the brane in relation to the origin.}%\footnote{Notice that $\lambda\to-\lambda$ agrees with $\beta\to -\beta$. Our best fit asks $\beta=-0.11$.}
\begin{equation}\label{v_eff}
V_{\rm eff}(\phi)=A_0(1-\beta\lambda\phi)^{\frac{1}{\beta}}+\frac12\sigma.
\end{equation}
Let us now consider two approaches in order to make a close connection with the Starobinsky inflaton potential. Firstly,  we consider the following arrangement 
\begin{equation}
V_{\rm eff}(\phi)=\frac12\sigma\left[1+2\frac{A_0}{\sigma}(1-\beta\lambda\phi)^{\frac{1}{\beta}}\right].
\end{equation}
For $A_0=-|A_0|$ with $|A_0|/\sigma\ll1$ we can make the following approximation 
\begin{equation}
V_{\rm eff}(\phi)=\frac12\sigma\left[1-\frac{|A_0|}{\sigma}(1-\beta\lambda\phi)^{\frac{1}{\beta}}\right]^2,
\end{equation}
which can still be recast in the form 
\begin{equation}
V(\phi)=V_0\left[1-\left(1-\beta\sqrt{\frac{2}{3}}\frac{\phi}{\rm M_{Pl}}\right)^{\frac{1}{\beta}}\right]^2,\label{eq:pot_ext}
\end{equation}
where $V_0=\frac12\sigma$, $\lambda=\sqrt{\frac{2}{3}}\frac{1}{\rm M_{Pl}}$, %$\phi$ is in Planck mass unit,
and we have absorbed the pre-factor $\frac{|A_0|}{\sigma}$ into the parenthesis, assuming that $\left(\frac{|A_0|}{\sigma}\right)^\beta\sim 1$. Notice that the expression (7) fully recovers the Starobinsky potential (\ref{eq:pot_sta}) for $\beta = 0$ whereas $\forall \beta \neq 0$ represents a generalized model that admits a wider range of solutions. 

%This is easily satisfied in the interval $\beta=-0.08\pm0.12$, say, for $(1/1000)^{-0.01}=1.07$, but hardly satisfied for the best fit $\beta=-0.11$, i.e., $(1/1000)^{-0.11}=2.13$, $(1/100)^{-0.11}=1.65$, $(1/10)^{-0.11}=1.28$, and so on. A way to circumvent this problem is transfer the smallness to $(1+\beta\lambda\phi)^{1/\beta}$ instead. In this case, the aforementioned pre-factor can be deliberately set to unit. The price to pay is that now we should be at the region of the potential around $\phi\gg1$ for $\beta<0$, which coincides with the flat region of the potential that is a nice region for inflationary phase purposes.

Let us now consider a second approach. We should recall that in brane cosmology there is a modification in the Friedmann equation induced in the brane as follows \cite{Binetruy:2000,Binetruy:2000a}
\begin{equation}
H^2=\frac{2}{3}\rho\left(1+\frac{\rho}{2\sigma}\right),
\end{equation}
where $\rho$ is the energy density and $\sigma$ is the brane tension. In the slow-roll regime we know that $\rho_{\rm eff}\sim V(\phi)$. Furthermore, at the limit $\rho/2\sigma\gg1$, the high energy limit, we find
\begin{eqnarray}
H^2=\frac{2}{3}\left(\frac{\rho^2}{2\sigma}\right), \nonumber\\
\simeq \frac{2}{3}\left(\frac{V_{\rm eff}(\phi)^2}{2\sigma}\right).
\end{eqnarray}
Now using the explicit form of the potential (\ref{v_eff}) we obtain
\begin{eqnarray}
H^2=\frac{2}{3}\frac{1}{2\sigma}\left[A_0(1-\beta\lambda\phi)^{\frac{1}{\beta}}+\frac12\sigma\right]^2 \nonumber\\
=\frac{2}{3}\frac{\sigma}{8}\left[1-(1-\beta\lambda\phi)^{\frac{1}{\beta}}\right]^2\sim\frac{2}{3}\rho_{\rm eff}
\end{eqnarray}
where in the second step we have assumed $A_0=-\frac12\sigma$. From above equation we finally read off the $\beta$-Starobinsky inflaton potential at high energy regime given by expression (7),
%\begin{eqnarray}
%V(\phi)=V_0\left[1-\left(1-\beta\sqrt{\frac{2}{3}}\phi\right)^{\frac{1}{\beta}}\right]^2,
%\end{eqnarray}
with $V_0\equiv\frac{1}{8}\sigma$ and $\lambda=\sqrt{\frac{2}{3}}\frac{1}{\rm M_{Pl}}$.%, and $\phi$ is in Planck mass unit.

\section{$\beta-$Starobinsky inflation}\label{betastarobinskyinflation}

%%%%%%%%%%%%%% wi %%%%%%%%%%%%%%%%%%%%%%%%
\begin{figure*}[!ht]
\centering
	\includegraphics[width=1\hsize]{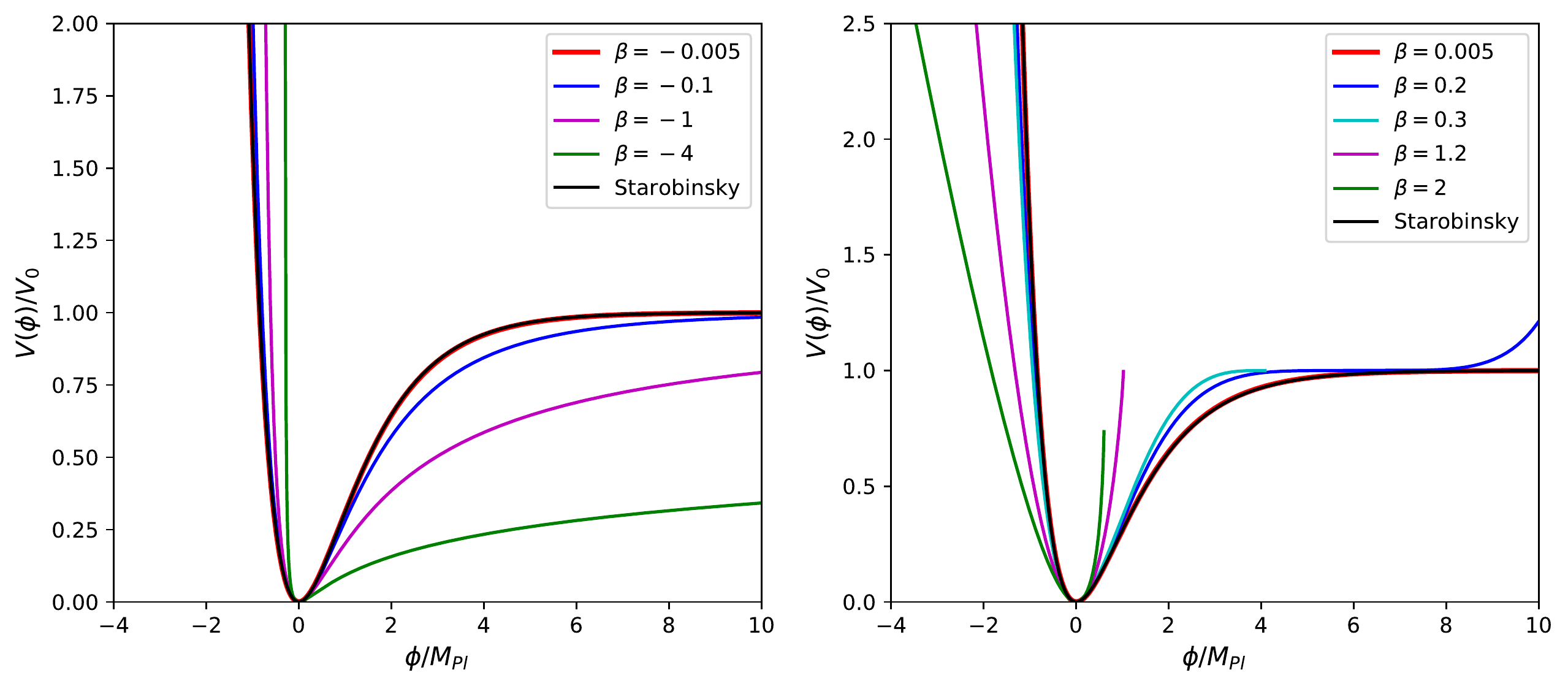}
	\caption{ {The potential of $\beta$-Starobinsky model for different values of $\beta$.}}
	\label{fig:pot}
\end{figure*}
%%%%%%%%%%%%%%%%%%%%%%%%%%%%%%%%%%%%

In this section, we discuss some theoretical predictions of the $\beta$-Starobinsky potential given by eq.~(7), discussed in the previous section. As we see, we can quantify how much this extended model deviates from the current best-fit inflationary model ($\beta = 0$) through the free parameter $\beta$.  Figure ~\eqref{fig:pot} shows the behaviour  of the potential (7) for some arbitrary values of $\beta$, and note that for both intervals $\beta< -4$ and  $\beta>0.3$ the potential behaviour differs significantly from the Starobinsky model. As expected, in the limit $\beta\rightarrow 0$ the potential \eqref{eq:pot_sta} is fully recovered. Furthermore, for $\beta<1.2$ one finds a large-field behaviour when $0<\phi<10$ whereas for values of $\beta>1.2$ we do not retrieve the large-field behaviour.

As is well known, one can characterize the slow-roll inflationary regime by parameters that depend on the field potential and its derivatives w.r.t the scalar field $\phi$ (denoted by the prime in the equations below). The slow-roll parameters for the model under consideration can be written as
\begin{eqnarray}
\epsilon(\phi) &=& M_{Pl}^2 \left[ \frac{V'(\phi)}{V(\phi)} \right]^2 = \frac{4}{3}\chi^{\frac{2}{\beta}-2}\left(1 - \chi^{\frac{1}{\beta}}\right)^{-2}, \label{cap6_eq_36} \\
\eta(\phi) &=& M_{Pl}^2 \frac{V''(\phi)}{V(\phi)}  = \frac{4}{3}\chi^{\frac{1}{\beta}-2}
{\left[ \beta - 1 - (\beta-2)\chi^{\frac{1}{\beta}} \right] \over 
\left(1 - \chi^{\frac{1}{\beta}}\right)^{2}},
\end{eqnarray}
where $\chi\equiv 1-\beta\sqrt{\frac{2}{3}}\frac{\phi}{M_{Pl}}$.

Inflation happens while $\epsilon, \eta \ll 1$ and the condition $\epsilon(\phi) = 1$ defines the value of the field $\phi$ when inflation ends, $\phi_{end}$. Since eq.~\eqref{cap6_eq_36} does not allow a direct inversion one needs to solve it numerically. We interpolated the points of $\beta$ and $\phi$ that satisfies the constraint $\epsilon(\phi) = 1$ with two polynomial fits of 12th order, which are solutions of eq.~\eqref{cap6_eq_36}: one is valid for $\phi_{end}>0$ and the other for $\phi_{end}<0$, and we call them solutions 1 and 2, respectively. Note that the Starobinsky model must be recovered when $\beta \rightarrow 0$, which happens only for the solution 1 (with $\phi_{end}\sim 0.94$). Hence, we discard the solution 2 as a viable extension of the Starobinsky model and, throughout this paper, we focus only on the investigation of the solution 1.

The potential amplitude, $V_0$, is obtained considering the primordial power spectrum of curvature perturbations, calculated  when the CMB mode exits from horizon at the scale $\phi_{*}$,
 \begin{eqnarray}
 P_R= \frac{1}{24\pi^2}\frac{V(\phi)}{\epsilon}\mid_{k=k_{*}}.
 \label{cap6_eq_43}
 \end{eqnarray}
The value of $P_R(k_{*})$ is determined by Planck normalization, i.e., $2.0933\times 10^{-9}$ for the pivot choice $k_{*}=0.05$Mpc$^{-1}$~\cite{Aghanim:2018eyx}. Combining eqs.~\eqref{eq:pot_ext} and~\eqref{cap6_eq_43}, and inverting for $V_0$, we obtain
\begin{eqnarray}
V_0 =  \frac{32\pi^2P_R\chi_{*}^{2/\beta -2}}{(1-\chi_{*}^{1/\beta})^4},
\label{cap6_eq_44}
\end{eqnarray}
where $\chi_{*}\equiv 1-\beta\sqrt{\frac{2}{3}}\frac{\phi_{*}}{M_{Pl}}$.

In order to find the value of $\phi_{*}$ we consider the number of $e$-folds since the CMB modes crossed the horizon until the end of inflation
%\begin{widetext}
\begin{equation}
N  = \int_{\phi_{end}}^{\phi}{\frac{d\phi}{\sqrt{2\epsilon}}} %= \int_{\phi_{end}}^{\phi}{\sqrt{\frac{3}{8}}\left[\left(1-\sqrt{\frac{2}{3}}\beta\frac{\phi}{M_{Pl}}\right)^{1-1/\beta}-1+\beta\sqrt{\frac{2}{3}}\frac{\phi}{M_{Pl}}\right]d\phi} 
%\nonumber\\
= \sqrt{\frac{3}{8}}M_{Pl}\left[- \left(\frac{\phi}{M_{Pl}}\right)  + \frac{\beta}{2}\sqrt{\frac{2}{3}}\left(\frac{\phi}{M_{Pl}}\right)^2  
%\nonumber\\
- \sqrt{\frac{3}{2}}\frac{1}{(2\beta - 1)} \left(1- \beta\sqrt{\frac{2}{3}}\frac{\phi}{M_{Pl}}\right)^{2-1/\beta}\right]_{\phi_{end}}^{\phi}.\label{eq:efolds}
%\end{align}
\end{equation}
%\begin{eqnarray}
%	N &=& \int_{\phi_{end}}^{\phi}{\frac{d\phi}{\sqrt{2\epsilon}}} %= \int_{\phi_{end}}^{\phi}{\sqrt{\frac{3}{8}}\left[\left(1-\sqrt{\frac{2}{3}}\beta\frac{\phi}{M_{Pl}}\right)^{1-1/\beta}-1+\beta\sqrt{\frac{2}{3}}\frac{\phi}{M_{Pl}}\right]d\phi}
%	 \nonumber \\
%	&=& \sqrt{\frac{3}{8}}M_{Pl}[- \left(\frac{\phi}{M_{Pl}}\right)  + \frac{\beta}{2}\sqrt{\frac{2}{3}}\left(\frac{\phi}{M_{Pl}}\right)^2  \nonumber 
%	 \end{eqnarray}
%	 \begin{eqnarray}
%	 ~~~- \sqrt{\frac{3}{2}}\frac{1}{(2\beta - 1)} \left(1- \beta\sqrt{\frac{2}{3}}\frac{\phi}{M_{Pl}}\right)^{2-1/\beta}|_{\phi_{end}}^{\phi}.
%	\label{cap6_eq_37}
%\end{eqnarray}
%\end{widetext}
with $\phi=\phi(N)$.

%%%%%%%%%%%%%% wi %%%%%%%%%%%%%%%%%%%%%%%%
\begin{figure*}[t]
\centering
	\includegraphics[width=0.5\hsize]{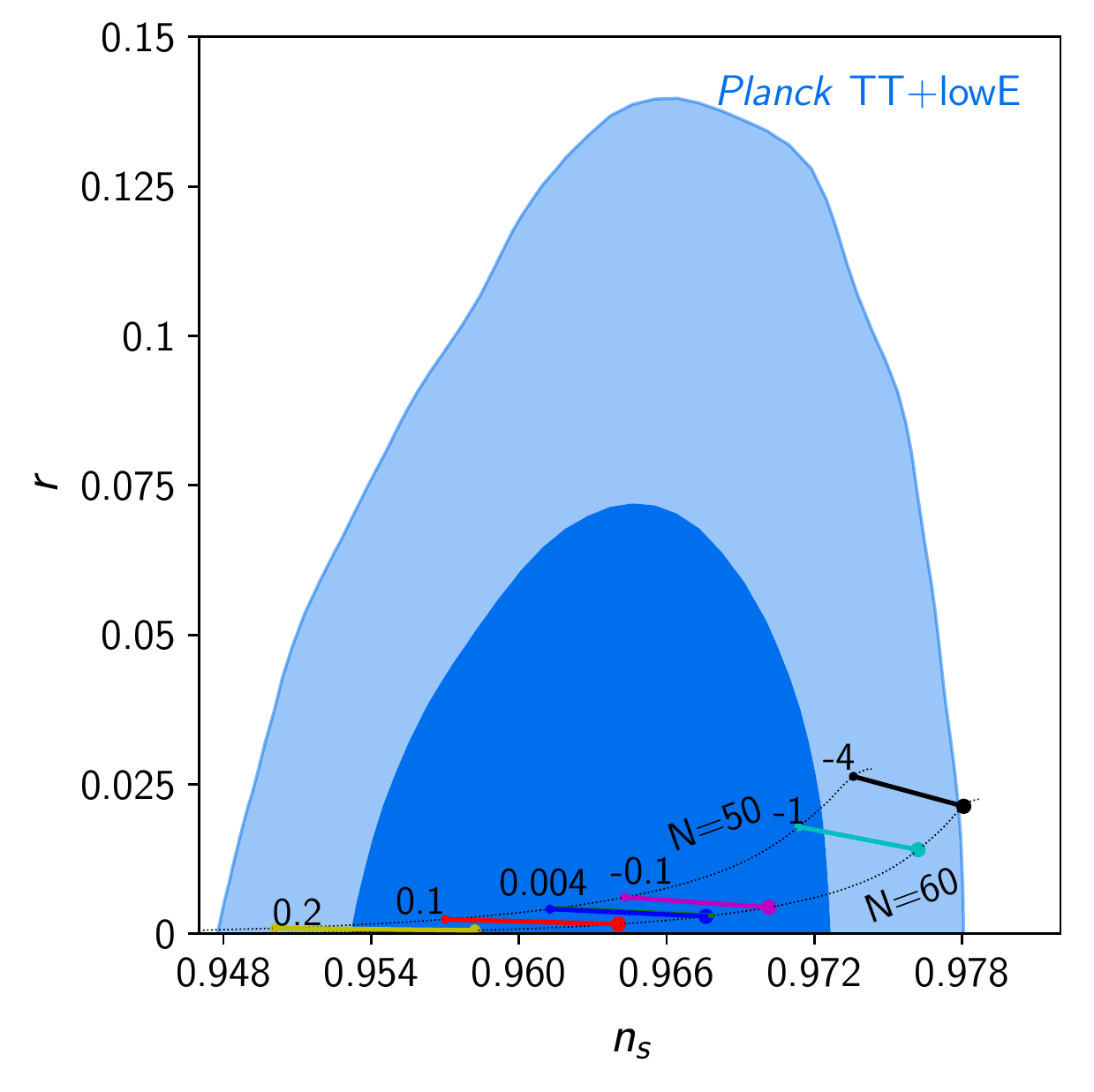}
	\caption{The $n_s - r$ plan for the values of $\beta$ satisfying eq.~\eqref{eq:efolds}, considering two values for the number of e-folds, $N=50$ and $N=60$. The contours are the $68\%$ and $95\%$ confidence level regions obtained from Planck (2018) CMB data using the pivot scale $k_{*}=0.05$Mpc$^{-1}$. }
	\label{fig:ns-r}
\end{figure*}
%%%%%%%%%%%%%%%%%%%%%%%%%%%%%%%%%%%%

Again, this expression can not be inverted and then we solve it numerically for $\phi_{*}(N)$. In the case in which the pivot scale crosses the Hubble horizon during inflation, we find the values of $\phi_{*}$ and $\beta$ for which $N=55$ is valid and interpolate with a polynomial fit for $\phi_{*}$.
Correspondingly, the value of the field in the beginning of inflation $\phi_{ini}$ is obtained when considering $N=70$ in eq.~\eqref{eq:efolds}. Similarly to $\phi_{end}$, the polynomial fits for $\phi_{*}$ and $\phi_{ini}$ are of 12th order and retrieve the Starobinsky model in the limit $\beta \rightarrow 0$. In addition, the slow-roll conditions are fully met for values of $-4<\beta<0.6$. 

Finally, the scalar spectral index, $n_s$, and the tensor-to-scalar ratio, $r$, are written as 
\begin{eqnarray}
n_s&=&1+\frac{8}{3}\frac{\chi^{\frac{1}{\beta}-2}}{\left(1-\chi^{\frac{1}{\beta}}\right)^2}\left[\beta\left(1-\chi^{\frac{1}{\beta}}\right)-1 - \chi^{\frac{1}{\beta}}\right], \\
r &=& \frac{64}{3}\chi^{\frac{2}{\beta}-2}\left(1-\chi^{\frac{1}{\beta}}\right)^{-2},\label{eq:ns}
\end{eqnarray}
and the consistency relation between $n_s$ and $r$ take the following form:
\begin{eqnarray}
r=8(n_s-1)\chi^{\frac{1}{\beta}}\left[ (\beta - 1) - \beta\chi^{\frac{1}{\beta}} \right].\label{eq:r}
\end{eqnarray}

The $n_s - r$ plane is shown in fig.~\eqref{fig:ns-r}. We display different values of $\beta$ satisfying the eq.~\eqref{eq:efolds} and consider two different numbers of e-folds, i.e., $N=50$ and $N=60$. The contours correspond to $68\%$ and $95\%$ confidence levels (C.L.) obtained from the most recent Planck CMB data~\cite{Aghanim:2018eyx}.  Notice that the values of $n_s$ and $r$ increase as the value of $\beta$ decreases. These results are not very restrictive because all the values predicted are within the $95\%$ region. The constrained values of $\beta$, $-4< \beta <0.2$, are consistent both with Planck results at 2$\sigma$ and with the slow-roll regime discussed earlier. Finally, it is also worth mentioning that even if the theoretical predictions of a given model are in agreement with the $n_s - r$ plane, it does not necessarily mean that it is a good model when compared with other inflationary scenarios~\cite{Campista:2017ovq}. Therefore, in what follows we will analyze the predictions of the power spectrum of temperature fluctuations and compare them with current CMB data.

\section{Method and analysis}\label{method}

The theoretical predictions of the $\beta$-Starobinsky model are calculated modifying the latest version of the Code for Anisotropies in the Microwave Background (CAMB)~\cite{camb}, to include the $\beta$ parameter, since in its standard realization it assumes a power-law parameterization for the primordial perturbation spectrum, $P_R=A_s(k/k_{*})^{n_s-1}$. In this context, we modify CAMB following the lines of the {\sc ModeCode} adapted for our primordial potential choice, in order to calculate the dynamic and perturbations of our model and then construct the theoretical predictions for the primordial power spectrum. 

\begin{figure}[]
\centering
	\includegraphics[width=0.5\hsize]{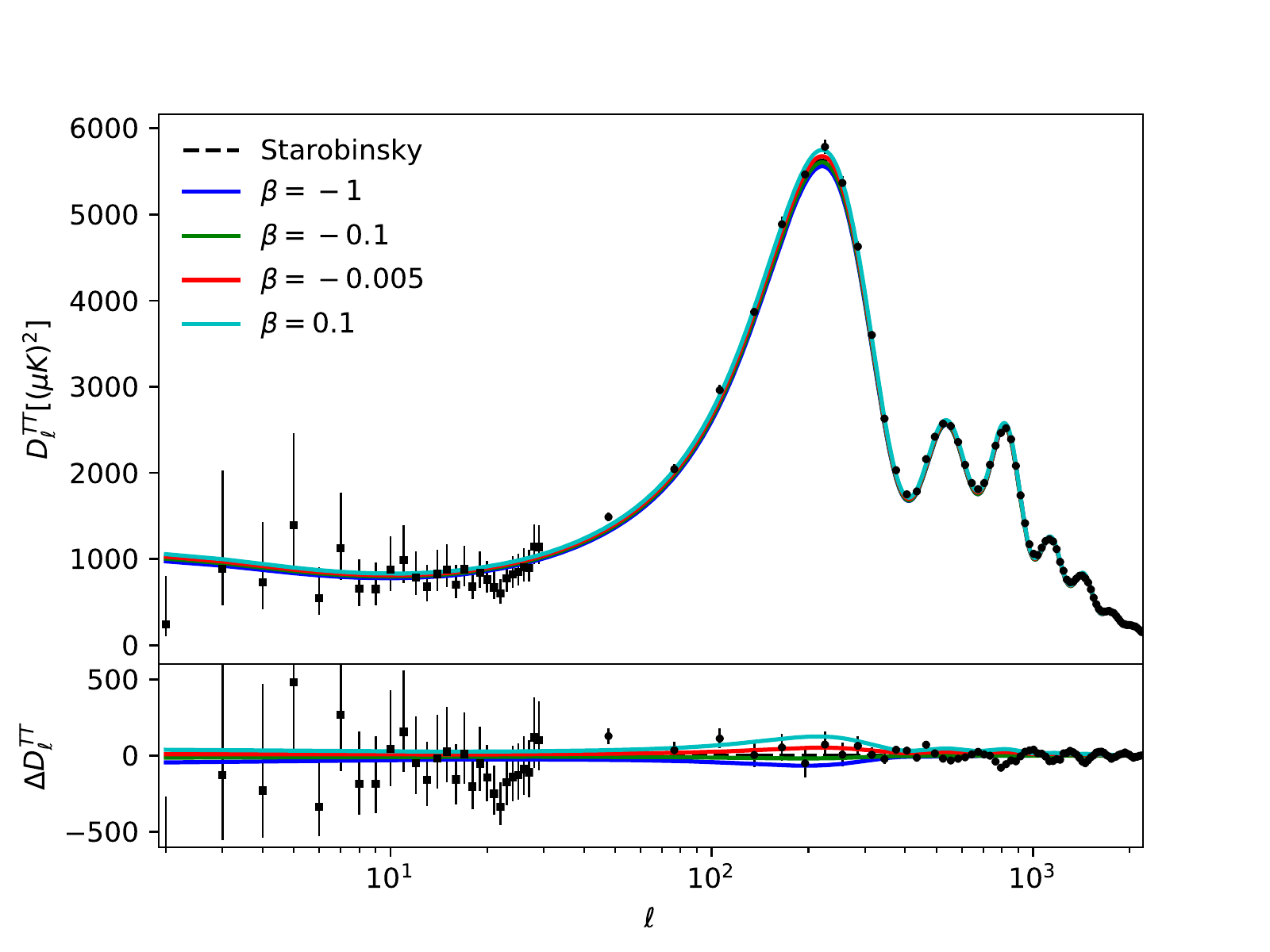}
	\caption{The theoretical predictions for the angular power spectra considering different values of $\beta$.} 
	\label{ps}
\end{figure}

{\sc ModeCode} calculates the spectrum of CMB temperature fluctuations solving numerically the equations of inflationary dynamics, namely the Friedmann and Klein-Gordon equations, as well as the Fourier components associated with curvature perturbations produced by the fluctuations of the scalar field $\phi$. These components are solution of the Mukhanov-Sasaki equations~\cite{Mukhanov:2005sc,weinberg2008cosmology}
\begin{eqnarray}
u''_k+\left(k^2-\frac{z''}{z}\right)u_k=0,
\end{eqnarray}
where $u\equiv -z\mathcal{R}$ and $z\equiv a\dot{\phi}/H$, and $a$, $H=\dot{a}/a$ and $\mathcal{R}$ are the scale factor, the Hubble parameter and the comoving curvature perturbations, respectively. The primordial power spectrum of curvature perturbations $\mathcal{P}(k)$ defined as  function of the vacuum expected value of $\mathcal{R}$ is
\begin{eqnarray}
<\mathcal{R}^{*}(k)\mathcal{R}(k')>=\frac{2\pi^2}{k^3}\delta^{3}(k-k')\mathcal{P}(k),
\end{eqnarray}
where $\delta$ is the Dirac delta function and the factor $\displaystyle2\pi^2/k^3$ is chosen  to obey the usual Fourier conventions. 
It then follows that $\mathcal{P_R}(k)$ is related with $u_k$ and $z$ via:
\begin{eqnarray}
\mathcal{P_R}(k)=\frac{k^3}{2\pi^2}\left|\frac{u_k}{z}\right| ^2.
\end{eqnarray}
Therefore, given the form of the inflaton potential $V(\phi)$, the dynamical equations are integrated to obtain $H$ and $\phi$ as function of time and then the solution $u_k$ for the mode $k$. Finally, it evaluates the spectrum of curvature perturbations when the mode crosses the horizon. 
The theoretical predictions of the $\beta-$Starobinsky potential are shown in fig.~\eqref{ps}. Note that the effect of the parameter $\beta$ is to slightly modify the amplitude of the temperature power spectrum.

%%%%%%%%%%%%%% wi %%%%%%%%%%%%%%%%%%%%%%%%
%%%%%%%%%%%%%%%%%%%%%%%%%%%%%%%%%%%%

%
%%%%%%%%%%%%%%%%%%%%%%%%
\begin{table}
\centering
\caption{Priors on the cosmological parameters considered in the analysis.}
{\begin{tabular}{|c|c|}
\hline 
Parameter & Prior Ranges \\ 
\hline
$\Omega_{b}h^{2}$ & $[0.005 : 0.1]$ \\ 
 
$\Omega_{c}h^{2}$ & $[0.001 : 0.99]$ \\ 

$\theta$ & $[0.5 : 10.0]$ \\ 
 
$\tau$ & $[0.01 : 0.8]$ \\ 
  
$\beta$ & $[-4 : 0.2]$ \\ 
\hline 
\end{tabular}\label{tab_priors}}
\end{table} 
%%%%%%%%%%%%%%%%%%%%%%%

In order to constrain the cosmological parameters associated with the $\beta$-Starobinsky model we perform an analysis using the latest version of {\sc CosmoMC} code~\cite{cosmomc}, necessary to explore the cosmological parameter space. In addition to the parameter $\beta$ we also vary the usual cosmological variables, namely, the baryon and the cold dark matter density, the ratio between the sound horizon and the angular diameter distance at decoupling, and the optical depth: $\left \{\Omega_bh^2~,~\Omega_ch^2~,~\theta~,~\tau\right \}$. We consider purely adiabatic initial conditions, fix the sum of neutrino masses to $0.06~eV$ and the universe curvature to zero, and also vary the nuisance foregrounds parameters~\cite{Aghanim:2015xee}. The  flat priors on the cosmological parameters used in our analysis are shown in 
table~\ref{tab_priors}. Moreover, the interval of values of the parameter $\beta$ is chosen from the considerations made in the previous sect., i.e., $-4< \beta <0.2$ (see e.g. fig.~\eqref{fig:ns-r}).

%data
We use the CMB data set from the latest Planck (2018) Collaboration release~\cite{Aghanim:2018eyx}, considering the high multipoles Planck temperature data from the 100-,143-, and 217-GHz half-mission T maps, and  the low multipoles data by the joint TT, EE, BB and TE likelihood, where EE and BB are the E- and B-mode CMB polarization power spectrum and TE is the cross-correlation temperature-polarization (hereafter PLC18). 
We also combine the CMB data with an extended background data sets composed of i) Baryon Acoustic Oscillations (BAO) from the 6dF Galaxy Survey (6dFGS)~\cite{bao1}, Sloan Digital Sky Survey (SDSS) DR7 Main Galaxy Sample galaxies~\cite{bao2}, BOSSgalaxy samples, LOWZ and CMASS~\cite{bao3} and ii) the tensor amplitude of B-mode polarization from 95, 150, and 220 GHz maps, which are the tightest and least model-dependent  constraints on the tensor amplitude coming from the Keck Array and BICEP2 Collaborations~\cite{bicep21,bicep22} analysis of the BICEP2/Keck field, in combination with Planck high-frequency maps to remove the polarized Galactic dust emission, used to constrain the parameters associated with the tensor spectrum (hereafter BK15).

\section{Results}\label{results}

\begin{figure*}[]
\centering
	\includegraphics[width=.75\hsize]{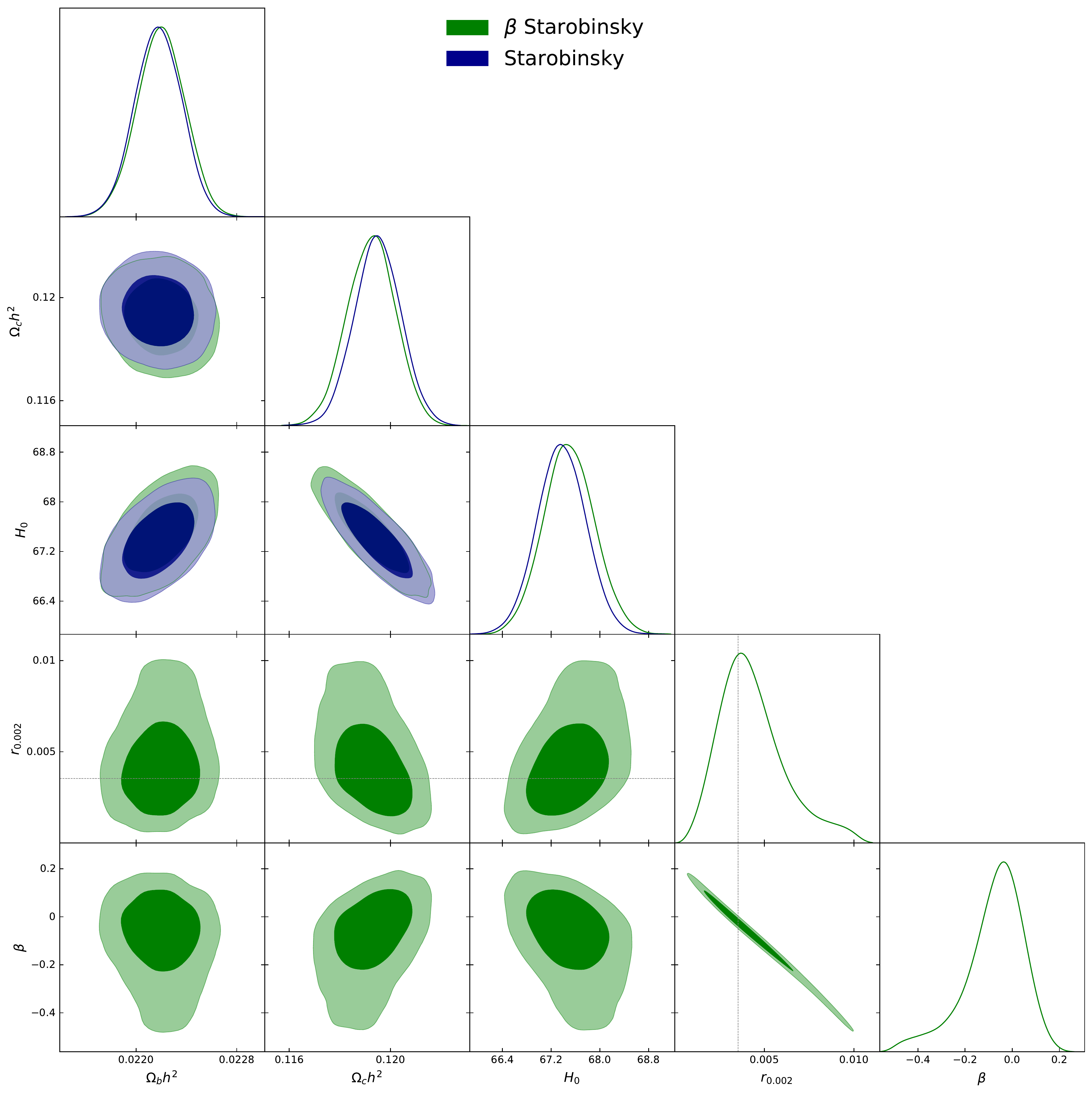}
	\caption{Two-dimensional probability distribution and one-dimensional probability distribution for the $\beta-$Starobinsky model (green contours) and the reference Starobinsky model (blue contours), both using the extended dataset (PLC18+BAO+BK15). The dotted lines indicate the predicted value for the tensor-to-scalar ratio $r$ for the standard Starobinsky model.}
	\label{fig:tri_plot}
\end{figure*}

The main results of our analysis are shown in table \ref{tab:Tabel_results_1}, where we summarize the constraints on the cosmological parameters of the Starobinsky and $\beta$-Starobinsky models obtained using the Planck 2018 likelihood combined with BAO and BK15 data.
We also show in fig.\eqref{fig:tri_plot} the  confidence intervals at $68\%$ and $95\%$ and the posterior probability distribution for the most interesting behaviours.  As we can see in the second column of table \ref{tab:Tabel_results_1}, all the primary and the derived cosmological parameters of $\beta$-starobinsky model are consistent within $1\sigma$ with the standard Starobinsky inflation. We found no evidence for a non-zero $\beta$ parameter, which is allowed to vary within the range $-0.08\pm 0.12$ (1$\sigma$). These results are also consistent with previous analyses~\cite{Renzi:2019ewp}, which have investigated a generalization of the Starobinsky inflation of the type $f(R)\propto R^{2p}$ and found $p \simeq 1$.

%%%%%%%%%%%%%%%%%%%%%%%%%%%%%%%%%%%%%%%%%%%%%%%%%%%%%%%
\begin{table*}
\centering
\caption{$68\%$ confidence limits and best fit values for the cosmological parameters. The  first  columns-block show the constraints on the parameters of the Starobinsky and $\beta-$Starobinsky models, using the extended data set, i.e. the joint PLC18+BAO+BKP15 data. 
The table is divided into two sections: the upper section shows the primary parameters, while in the lower part shows the derived ones and lastly the BIC values. The values indicated with ($^*$) are calculated for the pivot choice of $N=55$.}
\scalebox{1}{
{\begin{tabular}{|c|c|c|c|c|}
\hline
 &\multicolumn{2}{c|}{Starobinsky} &\multicolumn{2}{c|}{$\beta$-Starobinsky} \\
\hline
 {Parameter} & {mean} & {best fit} & {mean} & {best fit}\\
\hline
Primary & & & &\\
{$\Omega_b h^2$} 
& $0.02218 \pm 0.00018$
& $0.022276$
& $0.022198 \pm 0.00019$
& $0.022227 $	
\\
{$\Omega_{c} h^2$} 
& $0.1195 \pm 0.0009 $
& $0.11916 $ 
& $0.1192 \pm 0.0009$
& $0.11897$
\\
{$\theta$}
& $1.04092 \pm 0.00041 $
& $1.040603 $
& $1.04098 \pm 0.00041$
& $1.040920$
\\
{$\tau$}
& $0.0526 \pm 0.0028$
& $0.0547 $
& $0.0542 \pm 0.0044$
& $0.0523$
\\
{$\beta$}
& $-$
& $-$
& $-0.08 \pm 0.12$
& $-0.11 $
\\
\hline
\hline
Derived & & & &\\
{$H_{0}$}
& $67.37 \pm 0.40$
& $67.46 $
& $67.50 \pm 0.41$
& $67.60$
\\
{$\Omega_{m}$}
& $0.3136 \pm 0.005$
& $0.3122 $
& $0.3119 \pm 0.006$
& $0.3104$
\\
{$\Omega_{\Lambda}$}
& $0.6864 \pm 0.005 $
& $0.6878 $
& $0.6881 \pm 0.006$
& $0.6896$
\\
{$n_s$}
& $-$
& $0.9652^{*}$
& $-$
& $0.9675^{*}$
\\
{$r_{0.002}$}
& $ - $
& $ 0.0035^{*}$
& $0.0044 \pm 0.0018$
& $0.0048$
\\
\hline
\hline
$\Delta$BIC
&\multicolumn{2}{c|}{Reference} % Sta Ref 965.66
&\multicolumn{2}{c|}{Positive}%$2.7 $} % Sta 968.78 beta 968.24 
\\
\hline
\end{tabular} \label{tab:Tabel_results_1}}
}
\end{table*} 
%%%%%%%%%%%%%%%%%%%%%%

%
%%%%%%%%%%%%%% wi %%%%%%%%%%%%%%%%%%%%%%%%
%%%%%%%%%%%%%%%%%%%%%%%%%%%%%%%%%%%%

As discussed earlier, {\sc ModeCode} calculates the spectrum of CMB temperature fluctuations from the numerical solutions of inflationary dynamics, instead of a power-law parametrization in terms of the scalar amplitude $A_s$ and the spectral index $n_s$. This amounts to saying that the analyses we performed for both Starobinsky and $\beta$-Starobinsky models did not obtain direct constraints on those parameters, but we still can derive the spectral index through the eq.~\eqref{eq:ns} (see the derived $n_s$ values tagged with $'*'$ in table \ref{tab:Tabel_results_1}).
The constraints on tensor-to-scalar ratio $r$ for the $\beta$-Starobinsky model displayed in fig.~\eqref{fig:tri_plot} and table~\ref{tab:Tabel_results_1} show perfect agreement with Starobinsky inflation within $1\sigma$ for the theoretical value calculated here, with the upper limit reported in Planck 2018 release ($r<0.106$ at $95\%$ C.L.) and also with the lower limit of $r>0.0017$ at $95\%$ C.L. found by \cite{Renzi:2019ewp}.

Finally, in the last line of table~\ref{tab:Tabel_results_1} we also show the Bayesian Information Criterion (BIC), which consider only the point that maximizes the posterior probability distribution to compare the models, taking into account both the number of data points and the number of free parameters of the models under consideration. The BIC value is given by~\cite{schwarzbic}
$$
{\rm{BIC}} = -2\ln{{\cal L}(d|\theta)}+k\ln{N},
$$
where the number of free parameters are $k=4$ and $k=5$, for Starobinsky and $\beta-$Starobinsky models, respectively. We can rank the models using the $\Delta {\rm{BIC}}\equiv {\rm{BIC}}_i - {\rm{BIC}}_{ref}$ value, which represents the preference of the reference model over model $i$, with $\Delta {\rm{BIC}}\leq 2$, $2 < \Delta {\rm{BIC}} \leq 6$, $6 < \Delta {\rm{BIC}} \leq 10$ and $\Delta {\rm{BIC}} \geq 10$ meaning weak, positive, strong and very strong support for the reference model, respectively~\cite{bicscale}. We compare our generalized model with the original Starobinsky model and find $\Delta {\rm{BIC}} = 2.3$, which means that the Starobinsky model has a positive preference over the extended $\beta-$Starobinsky scenario. Therefore, even providing a good description of the data (for a small deviation of the Starobinsky model, $\Delta \beta = \pm0.12$), the generalized scenario is penalized by the presence of an extra parameter, that is, the data do not justify the extension of the Starobinsky model, preferring the minimum model. This result, therefore, reinforces the robustness of Starobinsky model to describe the primordial inflationary phase.

%
%%%%%%%%%%%%%% wi %%%%%%%%%%%%%%%%%%%%%%%%
\begin{figure}[]
\centering
	\includegraphics[width=0.5\hsize]{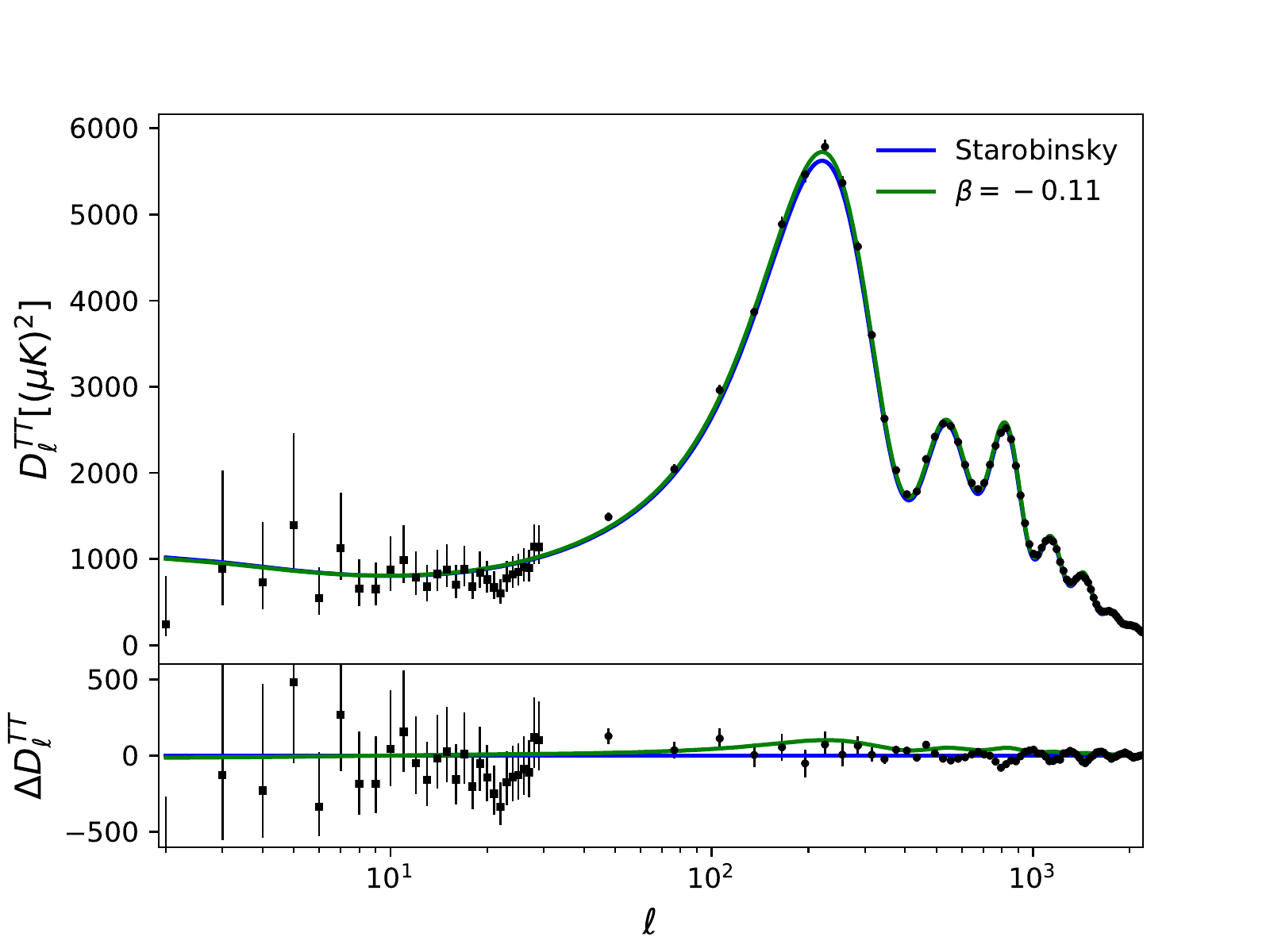}
	\caption{The best-fit angular power spectrum for the standard and $\beta-$Starobinsky models. The data points correspond to the latest release of Planck data~\cite{Aghanim:2018eyx} and the lower panel show the residuals with respect to the reference model (Starobinsky).}
	\label{fig:bestfit_TT}
\end{figure}
%%%%%%%%%%%%%%%%%%%%%%%%%%%%%%%%%%%%

\section{Conclusions}\label{conclusions}

{Quadratic f(R) theories can be understood as a sector of Lovelock gravity which may coincide with Einstein theory only in four-dimensions. At higher spacetime dimensions, however, it resembles string theory inspired models of gravity and, therefore, it seems natural to make a further connection of the Starobinsky model and its modifications as being part of the theories that can be found in fundamental setups such as in the realm of string/brane theories~\cite{Zwiebach:1985uq,Padmanabhan:2013xyr}.} In this paper, we investigated the robustness of the Starobinsky model by inserting it in the general framework of the $\beta$-Starobinsky potential derived from extra dimension physics. Moreover, we have analyzed this extension of the Starobinsky model motivated mainly by its remarkable observational success and by the works of \cite{Kallosh:2013xya,Kehagias:2013mya,Buchmuller:2013zfa,Benetti:2019smr,Capozziello:2015hra,vandeBruck:2015xpa,Renzi:2019ewp}, which show that the Starobinsky model can be retrieved by different approaches.

By using the most recent CMB measurements along with BAO and B-mode polarization data, we found that only small departures from the Starobinsky inflation is allowed within the range of $\beta=-0.08\pm0.12$ ($68\%$ C.L.), which implies a tensor-to-scalar ratio of $r_{0.002}=0.0044\pm 0.0018$ ($68\%$ C.L.). Such a result is in a good agreement with the currently available observational data, as shown in fig.~\eqref{fig:bestfit_TT}. 

As pointed out in \cite{Renzi:2019ewp}, the prediction of the Starobinsky model carries the uncertainties on $n_s$, thus $r$ could be due not to a real presence of tensor perturbations in Planck data but rather arising from the correlation between $r$ and $n_s$. Considering a generalisation of the type $R^{2p}$ in the Einstein-Hilbert action \eqref{eh}, these authors found limits on $r$ of the order of $r < 0.04$. On the other hand, the limits derived in our analysis show that the predicted value of the tensor-to-scalar ratio by the Starobinsky inflation differs by 0.5$\sigma$ or, equivalently, $\Delta r \sim 0.0009$ from the estimate obtained in the context of our extended scenario. Although such a small difference is not expected to be detectable by some future CMB experiments, such as LiteBIRD satellite~\cite{Suzuki:2018cuy} or Simons Observatory~\cite{Ade:2018sbj}, whose  sensitivities are $\Delta r \sim 0.001 - 0.002$~\cite{Ade:2018sbj,Suzuki:2018cuy}, it might be 
detected by the CMB-S4~\cite{Abazajian:2016yjj}, which is expected to reach the sensitivity of $\Delta r\sim 0.0006$.

Finally, despite the small deviations from the conventional Starobinsky model allowed by current observations, the BIC analysis indicates positive support for the Starobinsky model over the extended one. Therefore, the generalized potential proposed in this paper has allowed us to investigate the robustness of the Starobinsky inflation, and the statistical analysis performed has confirmed its remarkable success to describe current observational data. 

\begin{acknowledgement}

S. Santos da Costa acknowledges financial support from the Programa de Capacita\c{c}\~ao Institucional (PCI) do Observat\'orio Nacional/MCTI. M. Benetti acknowledges Istituto Nazionale di Fisica Nucleare (INFN), sezione di Napoli, iniziativa specifica QGSKY. R.M.P. Neves is supported by Coordena\c{c}\~{a}o de Aperfei\c{c}oamento de Pessoal de N\'ivel Superior (CAPES). F.A. Brito acknowledges support from Conselho Nacional de Desenvolvimento Cient\'{\i}fico e Tecnol\'ogico CNPq (Grant no. 312104/2018-9) and PRONEX/CNPq/FAPESQ-PB (Grant no. 165/2018). R. Silva acknowledges financial support from CNPq (Grant No. 303613/2015-7). J. Alcaniz is supported by CNPq (Grants no. 310790/2014-0 and 400471/2014-0) and Funda\c{c}\~ao de Amparo \`a Pesquisa do Estado do Rio de Janeiro FAPERJ (grant no. 233906). We also acknowledge the authors of the ModeCode (M. Mortonson, H. Peiris and R. Easther) and CosmoMC (A. Lewis) codes. This work was developed thanks to the High Performance Computing Center at the Universidade Federal do Rio Grande do Norte (NPAD/UFRN) and the Observat\'orio Nacional Data Center (DCON).
\end{acknowledgement}

%
% BibTeX users please use
% \bibliographystyle{}
% \bibliography{}

\begin{thebibliography}{}

\bibitem{Mukhanov:2005sc}
V.~Mukhanov,
`Physical Foundations of Cosmology, (Cambridge, 2005)

\bibitem{weinberg2008cosmology}
 Weinberg, S., Cosmology, (Oxford: OUP OXford, 2008)

\bibitem{Senatore:2016aui}
L.~Senatore,
Lectures on Inflation,
%doi:10.1142/9789813149441_0008
[arXiv:1609.00716 [hep-th]]

\bibitem{Planck2015} P.~A.~R.~Ade, {\it et al.} [Planck Collaboration], Astron.\ Astrophys.\  \textbf{594}, (2016) A13.

%\cite{Aghanim:2018eyx}
\bibitem{Aghanim:2018eyx} 
  N.~Aghanim {\it et al.} [Planck Collaboration],
  %Planck 2018 results. VI. Cosmological parameters,
  Astron.\ Astrophys.\  \textbf{641}, (2020) A6.
 % doi:10.1051/0004-6361/201833910
 % [arXiv:1807.06209 [astro-ph.CO]].
  %%CITATION = doi:10.1051/0004-6361/201833910;%%
  %3621 citations counted in INSPIRE as of 07 Oct 2020

\bibitem{Martin_2014} J. Martin, C. Ringeval, R. Trotta, V. Vennin, JCAP \textbf{03}, (2014) 039.

%\cite{Starobinsky:1980te}
\bibitem{Starobinsky:1980te} 
  A.~A.~Starobinsky,
  %``A New Type of Isotropic Cosmological Models Without Singularity,''
  Phys.\ Lett.\  \textbf{91B}, (1980) 99.
  [Adv.\ Ser.\ Astrophys.\ Cosmol.\  \textbf{3},  (1987) 130]
  %doi:10.1016/0370-2693(80)90670-X
  %%CITATION = doi:10.1016/0370-2693(80)90670-X;%%
  %4365 citations counted in INSPIRE as of 09 Dec 2019

%\cite{Ketov:2019toi}
\bibitem{Ketov:2019toi}
S.~V.~Ketov,
%``On the equivalence of Starobinsky and Higgs inflationary models in gravity and supergravity,''
J. Phys. A \textbf{53}, (2020) 084001.
%doi:10.1088/1751-8121/ab6a33
%[arXiv:1911.01008 [hep-th]].
%6 citations counted in INSPIRE as of 20 Oct 2020
%

%\cite{Linde:2011nh}
\bibitem{Linde:2011nh} 
  A.~Linde, M.~Noorbala and A.~Westphal,
  %``Observational consequences of chaotic inflation with nonminimal coupling to gravity,''
  JCAP \textbf{1103}, (2011) 013.
  %doi:10.1088/1475-7516/2011/03/013
  %[arXiv:1101.2652 [hep-th]].
  %%CITATION = doi:10.1088/1475-7516/2011/03/013;%%
  %88 citations counted in INSPIRE as of 06 Nov 2019

%\cite{Kallosh:2013xya}
\bibitem{Kallosh:2013xya} 
  R.~Kallosh and A.~Linde,
  %``Superconformal generalizations of the Starobinsky model,''
  JCAP \textbf{1306}, (2013) 028.
  %doi:10.1088/1475-7516/2013/06/028
  %[arXiv:1306.3214 [hep-th]].
  %%CITATION = doi:10.1088/1475-7516/2013/06/028;%%
  %211 citations counted in INSPIRE as of 07 Oct 2019
 
\bibitem{Kehagias:2013mya}
A.~Kehagias, A.~Moradinezhad Dizgah and A.~Riotto,
%``Remarks on the Starobinsky model of inflation and its descendants,''
Phys. Rev. D \textbf{89} (2014)  043527.
%doi:10.1103/PhysRevD.89.043527
  
%\cite{Buchmuller:2013zfa}
\bibitem{Buchmuller:2013zfa}
W.~Buchmuller, V.~Domcke and K.~Kamada,
%``The Starobinsky Model from Superconformal D-Term Inflation,''
Phys. Lett. B \textbf{726}, (2013) 467-470.
%doi:10.1016/j.physletb.2013.08.042
%[arXiv:1306.3471 [hep-th]].
%84 citations counted in INSPIRE as of 17 Jul 2020
 
%\cite{Benetti:2019smr}
\bibitem{Benetti:2019smr}
M.~Benetti, S.~Capozziello and L.~L.~Graef,
%``Swampland conjecture in $f(R)$ gravity by the Noether Symmetry Approach,''
Phys. Rev. D \textbf{100} (2019), 084013. 
%doi:10.1103/PhysRevD.100.084013
%[arXiv:1905.05654 [gr-qc]].
%9 citations counted in INSPIRE as of 10 Nov 2020

%\cite{Capozziello:2015hra}
\bibitem{Capozziello:2015hra}
S.~Capozziello, G.~Gionti, S.J. and D.~Vernieri,
%``String duality transformations in $f(R)$ gravity from Noether symmetry approach,''
JCAP \textbf{01}, (2016) 015.
%doi:10.1088/1475-7516/2016/01/015
%[arXiv:1508.00441 [gr-qc]].
%19 citations counted in INSPIRE as of 10 Nov 2020
  
\bibitem{Abazajian:2016yjj}
K.~N.~Abazajian, \textit{et al.} [CMB-S4],
%``CMB-S4 Science Book, First Edition,''
[arXiv:1610.02743 [astro-ph.CO]]


%\cite{Suzuki:2018cuy}
\bibitem{Suzuki:2018cuy} 
  A.~Suzuki, {\it et al.},
  %``The LiteBIRD Satellite Mission - Sub-Kelvin Instrument,''
  J.\ Low.\ Temp.\ Phys.\  \textbf{193}, (2018) 1048.
%  doi:10.1007/s10909-018-1947-7
 % [arXiv:1801.06987 [astro-ph.IM]].
  %%CITATION = doi:10.1007/s10909-018-1947-7;%%
  %49 citations counted in INSPIRE as of 06 Nov 2019

\bibitem{Ade:2018sbj}
P.~Ade, \textit{et al.} [Simons Observatory],
%``The Simons Observatory: Science goals and forecasts,''
JCAP \textbf{02},  (2019) 056.
%doi:10.1088/1475-7516/2019/02/056
%[arXiv:1808.07445 [astro-ph.CO]].

%\cite{vandeBruck:2015xpa}
\bibitem{vandeBruck:2015xpa} 
  C.~van de Bruck and L.~E.~Paduraru,
  %``Simplest extension of Starobinsky inflation,''
  Phys.\ Rev.\ D \textbf{92}, (2015) 083513.
  %doi:10.1103/PhysRevD.92.083513
  %[arXiv:1505.01727 [hep-th]].
  %%CITATION = doi:10.1103/PhysRevD.92.083513;%%
  %23 citations counted in INSPIRE as of 07 Oct 2019
  
%\cite{Renzi:2019ewp}
\bibitem{Renzi:2019ewp}
F.~Renzi, M.~Shokri and A.~Melchiorri,
%``What is the amplitude of the gravitational waves background expected in the Starobinski model?,''
Phys. Dark Univ. \textbf{27}, (2020) 100450.
%doi:10.1016/j.dark.2019.100450
%[arXiv:1909.08014 [astro-ph.CO]].
%0 citations counted in INSPIRE as of 13 Jul 2020

%\cite{Sebastiani:2013eqa}
\bibitem{Sebastiani:2013eqa}
L.~Sebastiani, G.~Cognola, R.~Myrzakulov, S.~D.~Odintsov and S.~Zerbini,
%``Nearly Starobinsky inflation from modified gravity,''
Phys. Rev. D \textbf{89} (2014) 023518.
%doi:10.1103/PhysRevD.89.023518
%[arXiv:1311.0744 [gr-qc]].
%115 citations counted in INSPIRE as of 02 Oct 2020

%\cite{Myrzakulov:2014hca}
\bibitem{Myrzakulov:2014hca}
R.~Myrzakulov, S.~Odintsov and L.~Sebastiani,
%``Inflationary universe from higher-derivative quantum gravity,''
Phys. Rev. D \textbf{91} (2015) 083529.
%doi:10.1103/PhysRevD.91.083529
%[arXiv:1412.1073 [gr-qc]].
%53 citations counted in INSPIRE as of 02 Oct 2020


\bibitem{Alcaniz:2006nu} 
 J.~S.~Alcaniz and F.~C.~Carvalho,
%  %``Beta-exponential inflation,''
EPL \textbf{79}, (2007) 39001.
%  doi:10.1209/0295-5075/79/39001
%[astro-ph/0612279].

%\cite{Santos:2017alg}
\bibitem{Santos:2017alg} 
  M.~A.~Santos, M.~Benetti, J.~Alcaniz, F.~A.~Brito and R.~Silva,
  %``CMB constraints on $\beta$-exponential inflationary models,''
  JCAP \textbf{1803}, (2018) 023.
  %doi:10.1088/1475-7516/2018/03/023
  % [arXiv:1710.09808 [astro-ph.CO]].
  %%CITATION = doi:10.1088/1475-7516/2018/03/023;%%
  %10 citations counted in INSPIRE as of 28 Nov 2019
  
  \bibitem{Gron:2018rtj}
$\O$.~Gr$\o$n,
%``Predictions of Spectral Parameters by Several Inflationary Universe Models in Light of the Planck Results,''
Universe \textbf{4}, (2018) 15.

\bibitem{Binetruy:2000}
  P.~Binetruy, C.~Deffayet and D.~Langlois,
  Nul. Phys. B \textbf{565}, (2000) 269.
  
\bibitem{Binetruy:2000a}
 P.~Binetruy, C.~Deffayet, U.~Ellwanger and D.~Langlois,
 Phys. Lett. B \textbf{477}, (2000) 285.
  
  
%\cite{Campista:2017ovq}
\bibitem{Campista:2017ovq} 
  M.~Campista, M.~Benetti and J.~Alcaniz,
  %``Testing non-minimally coupled inflation with CMB data: a Bayesian analysis,''
  JCAP \textbf{1709}, (2017) 010.
  %doi:10.1088/1475-7516/2017/09/010
 % [arXiv:1705.08877 [astro-ph.CO]].
  %%CITATION = doi:10.1088/1475-7516/2017/09/010;%%
  %3 citations counted in INSPIRE as of 06 Nov 2019	
  

\bibitem{camb}
A. Lewis, A. Challinor and A. Lasenby,
% ``Efficient computation of CMB anisotropies in closed FRW models'',
 Astrophys. J.  \textbf{538}, (2000)  473.


\bibitem{cosmomc} 
A. Lewis and S. Bridle,
% ``Cosmological parameters from CMB and other data: A Monte Carlo approach'',
 Phys. Rev. D  \textbf{66}, (2002) 103511.

%\cite{Aghanim:2015xee}
\bibitem{Aghanim:2015xee} 
  N.~Aghanim {\it et al.} [Planck Collaboration],
  %``Planck 2015 results. XI. CMB power spectra, likelihoods, and robustness of parameters,''
  Astron.\ Astrophys.\  \textbf{594},(2016)  A11.
  %doi:10.1051/0004-6361/201526926
  %[arXiv:1507.02704 [astro-ph.CO]].
  %%CITATION = doi:10.1051/0004-6361/201526926;%%
  %673 citations counted in INSPIRE as of 06 Nov 2019

\bibitem{bao1}
F.  Beutler,   \textit{et al.}
%C.  Blake,  M.  Colless,  D.  Heath  Jones,L. Staveley-Smith, L. Campbell, Q. Parker, W. Saunders,and F. Watson,
Mon. Not. R. Astron. Soc.  \textbf{416}, (2011) 3017.

\bibitem{bao2}
A. J. Ross,  \textit{et al.}
% L. Samushia, C. Howlett, W. J. Percival, A.Burden, and M. Manera,
Mon. Not. R. Astron. Soc. \textbf{449}, (2015) 835.

\bibitem{bao3}
L. Anderson, et al.(BOSS Collaboration), Mon. Not. R.Astron. Soc.  \textbf{441} (2014) 24.

\bibitem{bicep21}
P. A. R. Ade, {\it et al.} (BICEP2 and Planck Collaborations), Phys. Rev. Lett.  \textbf{114}, (2015) 101301.

\bibitem{bicep22}
P. A. R. Ade, {\it et al.} (BICEP2 and Keck Array Collaborations), Phys. Rev. Lett.  \textbf{116}, (2016)  031302.

%\cite{schwarzbic}
\bibitem{schwarzbic}
Schwarz, G., % Estimating the Dimension of a Model.
 Ann. Statist. \textbf{6},  (1978) 2.%doi:10.1214/aos/1176344136. %https://projecteuclid.org/euclid.aos/1176344136
%@article{schwarz1978,
%author = "Schwarz, Gideon",
%doi = "10.1214/aos/1176344136",
%fjournal = "Annals of Statistics",
%journal = "Ann. Statist.",
%month = "03",
%number = "2",
%pages = "461--464",
%publisher = "The Institute of Mathematical Statistics",
%title = "Estimating the Dimension of a Model",
%url = "https://doi.org/10.1214/aos/1176344136",
%volume = "6",
%year = "1978"
%}

%\cite{bicscale}
\bibitem{bicscale}
{R. Kass and A. Raftery},
% Bayes factors,
J.\ Am.\ Statist.\ Assoc.\ \textbf{90}, (1995) 773.

%\cite{Zwiebach:1985uq}
\bibitem{Zwiebach:1985uq}
B.~Zwiebach,
%``Curvature Squared Terms and String Theories,''
Phys. Lett. B \textbf{156}, (1985), 315-317.
%doi:10.1016/0370-2693(85)91616-8
%1071 citations counted in INSPIRE as of 12 Nov 2020

%\cite{Padmanabhan:2013xyr}
\bibitem{Padmanabhan:2013xyr}
T.~Padmanabhan and D.~Kothawala,
%``Lanczos-Lovelock models of gravity,''
Phys. Rept. \textbf{531} (2013), 115-171.
%doi:10.1016/j.physrep.2013.05.007
%[arXiv:1302.2151 [gr-qc]].
%139 citations counted in INSPIRE as of 12 Nov 2020

%Author, \textit{Book title} (Publisher, place year) page numbers
\end{thebibliography}

\end{document}